\def\appropto{\mathrel{\vcenter{
  \offinterlineskip\halign{\hfil$##$\cr
    \propto\cr\noalign{\kern2pt}\sim\cr\noalign{\kern-2pt}}}}}

\def\keV{\rm keV}

\documentclass[iop]{emulateapj}
\begin{document}

\slugcomment{\apj , submitted 4 Aug 2017 (Printed \today)}
\title{A General Precipitation-Limited $L_{\rm X}$--$T$--$R$ Relation Among Early-Type Galaxies}
\author{G. Mark Voit\altaffilmark{1,2}, C. P. Ma\altaffilmark{3}, J. Greene\altaffilmark{4}, A. Goulding\altaffilmark{4}, V. Pandya\altaffilmark{5}, M. Donahue\altaffilmark{1}, M. Sun\altaffilmark{6}}  
\altaffiltext{1}{Department of Physics and Astronomy,
                 Michigan State University,
                 East Lansing, MI 48824} 
\altaffiltext{2}{voit@pa.msu.edu}
\altaffiltext{3}{Departments of Astronomy and Physics,
                      University of California, Berkeley,
                      Berkeley, CA}
\altaffiltext{4}{Department of Astrophysical Sciences,
                      Princeton University,
                      Princeton, NJ}
\altaffiltext{5}{UCO/Lick Observatory, 
                      Department of Astronomy and Astrophysics
                      University of California, 
                      Santa Cruz, CA}
\altaffiltext{6}{Department of Physics,
                      University of Alabama Huntsville,
                      Huntsville, AL}
%\altaffiltext{6}{Other places}

\begin{abstract}
The relation between X-ray luminosity ($L_X$) and ambient gas temperature ($T$) among massive galactic systems is an important cornerstone of both observational cosmology and galaxy-evolution modeling.  In the most massive galaxy clusters, the relation is determined primarily by cosmological structure formation.  In less massive systems, it primarily reflects the feedback response to radiative cooling of circumgalactic gas.  Here we present a simple but powerful model for the $L_X$--$T$ relation as a function of physical aperture $R$ within which those measurements are made.  The model is based on the precipitation framework for AGN feedback and assumes that the circumgalactic medium is precipitation-regulated at small radii and limited by cosmological structure formation at large radii.  We compare this model with many different data sets and show that it successfully reproduces the slope and upper envelope of the $L_X$--$T$--$R$ relation over the temperature range from $\sim 0.2$~keV through $\gtrsim 10$~keV.  Our findings strongly suggest that the feedback mechanisms responsible for regulating star formation in individual massive galaxies have much in common with the precipitation-triggered feedback that appears to regulate galaxy-cluster cores.
\end{abstract}

\vspace*{2em}

\section{Introduction}

\setcounter{footnote}{0}

Massive galactic systems have long been known to have X-ray luminosities ($L_{\rm X}$) that are closely related to the temperature ($T$) of the diffuse ambient gas that fills them \citep{Sarazin_1986RvMP...58....1S,Mulchaey_2000ARA&A..38..289M,MathewsBrighenti2003ARAA..41..191M}.  In clusters of galaxies ($kT \gtrsim 2$~keV) the observed relation is $L_{\rm X} \propto T^{\zeta}$ with $\zeta \approx 2.5$--3 \citep[e.g.,][]{Mushotzky_1984PhST....7..157M,Markevitch_1998_LX-T,Pratt_2009_REXCESS_LX-T,Maughan_2012_LX-T, Giles_2016_XXL_LX-T}.  Groups of galaxies (1~keV~$\lesssim kT \lesssim 2$~keV) exhibit a slightly steeper power-law relationship with $\zeta \approx 3$--$3.5$ \citep[e.g.,][]{OsmondPonman_2004MNRAS.350.1511O, Sun+09}.  Among early-type galaxies with $kT \lesssim 1$~keV, the observed relationship steepens further to $\zeta \approx 4.5$ \citep[e.g.,][]{O'Sullivan_2003MNRAS.340.1375O,Boroson_2011ApJ...729...12B,KimFabbiano_2015ApJ...812..127K,Goulding_2016ApJ...826..167G}.

The $L_{\rm X}$--$T$ relationship steepens toward lower temperatures because non-gravitational processes such as radiative cooling and energetic feedback have progressively greater effects on the structure of the ambient medium as the depth of the confining gravitational potential declines \citep[e.g.,][]{Kaiser_1991ApJ...383..104K,EvrardHenry_1991ApJ...383...95E,pcn99,Voit_2005RvMP...77..207V}.   In the absence of cooling and feedback, gravitational structure formation would have produced a nearly self-similar family of objects with $L_{\rm X} \propto T^{3/2} \Lambda(T)$, where $\Lambda$ is the usual cooling function for optically thin gas in collisional ionization equilibrium.   However, ambient gas in a galactic system radiates an amount of energy similar to its thermal energy in a time $t_{\rm cool} = 3nkT  / 2 n_e n_i \Lambda(T)$, where $n_e$, $n_i$, and $n$ are the electron density, ion density, and total number density, respectively.   Near the center of a typical galactic system, this cooling time is less than the age of the universe.  Consequently, radiative cooling and whatever feedback it triggers inevitably modify the density distribution of the ambient medium and the resulting $L_{\rm X}$--$T$ relation, with more pronounced effects at lower temperatures \citep[e.g.,][]{vb01,Voit+02}.

Numerical simulations of enormous sophistication and complexity have been devoted to modeling the interplay between cooling and feedback in massive galaxies \citep[e.g.,][]{Vogelsberger_Illustris_2014Natur.509..177V,Hopkins_FIRE_2014MNRAS.445..581H,Schaye_EAGLE_2015MNRAS.446..521S}, but observations are increasingly indicating that a phenomenological principle of surprising simplicity emerges from those complex interactions:  Feedback triggered by production of cold gas clouds appears to limit $t_{\rm cool}$ at a given radius $R$ to be no less than $\approx 10$ times the freefall time $t_{\rm ff} = (2R/g)^{1/2}$, where $g$ is the gravitational acceleration at $R$ \citep[e.g.,][]{Sharma+2012MNRAS.427.1219S}.  This lower limit on $t_{\rm cool}(R)$ is observed among both central cluster galaxies \citep{Voit_2015Natur.519..203V,Hogan_2017_tctff} and other massive elliptical galaxies \citep{Voit+2015ApJ...803L..21V}.  It is also observed in numerical simulations in which chaotic accretion of cold gas clouds towards the central black hole fuels strong bipolar outflows \citep[e.g.,][]{Gaspari+2012ApJ...746...94G,Li_2015ApJ...811...73L,Meece_2017ApJ...841..133M}.

Initially, the theoretical rationale for a feedback-enforced lower limit at $t_{\rm cool} / t_{\rm ff} \approx 10$ was framed in terms of thermal instability \citep{McCourt+2012MNRAS.419.3319M,Sharma_2012MNRAS.420.3174S}.  However, numerical simulations have since shown that uplift of low-entropy ambient gas is at least as important to production of cold clouds.  As in a thunderstorm, the adiabatic cooling that occurs during uplift induces a phase transition \citep[e.g.,][]{Revaz_2008A&A...477L..33R,LiBryan2014ApJ...789..153L,McNamara_2016arXiv160404629M}.  In this case, uplift induces condensation of gas clouds much cooler and denser than the ambient gas.  Modest amounts of vertical mixing therefore lead to development of multiphase structure in gas that initially has $t_{\rm cool}/t_{\rm ff} \lesssim 10$.  Newly-formed cold clouds then rain toward the center of the galaxy, and the analogy with terrestrial weather has led us to call the overall process ``precipitation."   It is self-regulating because feedback fueled by the condensates heats the ambient medium until $\min (t_{\rm cool} / t_{\rm ff}) \gtrsim 10$, and that rise in $t_{\rm cool} / t_{\rm ff}$ inhibits further precipitation, for reasons discussed extensively in \citet{Voit_2017_BigPaper}.

This paper presents evidence showing that the ``precipitation limit'' observed at $t_{\rm cool} / t_{\rm ff} \approx 10$ in massive galactic systems extends down to Milky Way scales.  In Section~\ref{sec-LX(R)}, we compute the maximum X-ray luminosity that can come from within radius $R$ in a system with $t_{\rm cool} \gtrsim 10 t_{\rm ff}$ at all radii.  In Section~\ref{sec-LXTR} we compare this limit to observations of massive galactic systems ranging from galaxy clusters down to individual early-type galaxies from the ATLAS$^{\rm 3D}$ \citep{Cappellari_2011_ATLAS3D} and MASSIVE \citep{Ma_2014_MASSIVE} samples.  This comparison shows that the upper envelope of $L_X$ within $R$ among these systems matches the precipitation limit over nearly 7 orders of magnitude.  Section~\ref{sec-LX_T_R500} then builds the precipitation limit into a more general model for the $L_X$--$T$--$R$ relation and compares it with some representative data.  Section~\ref{sec-summary} summarizes our results.

\section{A Precipitation Limit on $L_{\rm X}(R)$}
\label{sec-LX(R)}

In a spherically symmetric system, the X-ray luminosity from within radius $R$ is
\begin{equation}
  L_{\rm X}(R) \: = \: \int_0^{R}  4 \pi r^2  \, n_e n_i  \Lambda(T) \, dr 
   \; \; .
\end{equation}
According to the precipitation framework, feedback triggered by condensation enforces $t_{\rm cool} \gtrsim 10 t_{\rm ff}$ and places an upper bound on the electron density of the ambient gas:
\begin{equation}
  n_e \: \lesssim \: \frac {3 kT} {10 \, t_{\rm ff} \Lambda(T)} \left( \frac {n} {2 n_i} \right) 
    \; \; .
    \label{eq-max_ne}
\end{equation}
Assuming an isothermal potential ($t_{\rm ff} \propto R$) in which the ambient gas temperature remains approximately constant with radius, we obtain from this constraint the following upper limit on the X-ray luminosity coming from within $R$:
\begin{equation}
  L_{\rm X}(R) \: \lesssim \:  \frac {9 \pi} {25} \frac {(kT)^2}  {\Lambda(T)} \, \sigma_v^2 \, R
  \; \; .
   \label{eq-LXTRsigv}
\end{equation} 
Here, $\sigma_v = (gR)^{1/2} $ is the line-of-sight component of an isotropic velocity dispersion corresponding to the potential-well depth, and we have dropped the insignificant $n^2 / 4 n_e n_i$ factor for simplicity.

Expressing $\sigma_v$ in terms of the parameter $\beta \equiv \mu m_p \sigma_v^2 / kT$ reduces this upper limit on X-ray luminosity to a function of $T$ and $R$:
\begin{equation}
  L_{\rm X}(R) \: \lesssim \:   \frac {9 \pi} {25} \frac {\beta} {\mu m_p} \frac {(kT)^3}  { \Lambda(T)}  \, R
  \; \; .
   \label{eq-LXTRbeta}
\end{equation} 
In systems that are close to hydrostatic equilibrium, the value of $\beta$ reflects the power-law slope of the radial gas pressure gradient, because $d \ln P / d \ln r \approx - 2 \beta$.  A nearly isothermal and hydrostatic system at the precipitation limit therefore has $P(r) \appropto n_e(r) \appropto r^{-1}$ and $\beta \approx 0.5$.

The factor of $\Lambda(T)$ in the denominators of equations (\ref{eq-max_ne}), (\ref{eq-LXTRsigv}), and (\ref{eq-LXTRbeta}) is a novel feature of precipitation-limited systems.  It ends up in the denominator because raising $\Lambda(T)$ increases the ambient medium's capacity for cooling and condensation.  The resulting energetic feedback therefore does not diminish until it drives down the ambient gas density far enough to limit condensation.  Within a fixed radius $R$, the limiting $L_X$--$T$ relation in the free-free cooling regime ($kT \gtrsim 2$~keV) is characterized by $\zeta \approx 2.5$.  At lower temperatures the relation becomes steeper, as emission-line cooling starts to dominate.  
These features agree well with observations of the $L_X$--$T$ relation and provide a particularly natural explanation for the observed steepening below 2~keV.  However, a proper comparison between theory and observation becomes more difficult at lower temperatures because cooler systems have substantially lower X-ray surface brightnesses.   This feature complicates the task of measuring $L_X(R)$ at fixed $R$ over a broad range in system mass. So instead, we will proceed to compare observations of $L_X(R)$ to the predicted luminosity limit over the ranges in $R$ for which good measurements are available.

\section{Comparison with Observations}
\label{sec-LXTR}

% ----------------------------------------
\begin{figure*}[t]
\begin{center}
\includegraphics[width=7in, trim = 0.1in 0.1in 0.0in 0.0in]{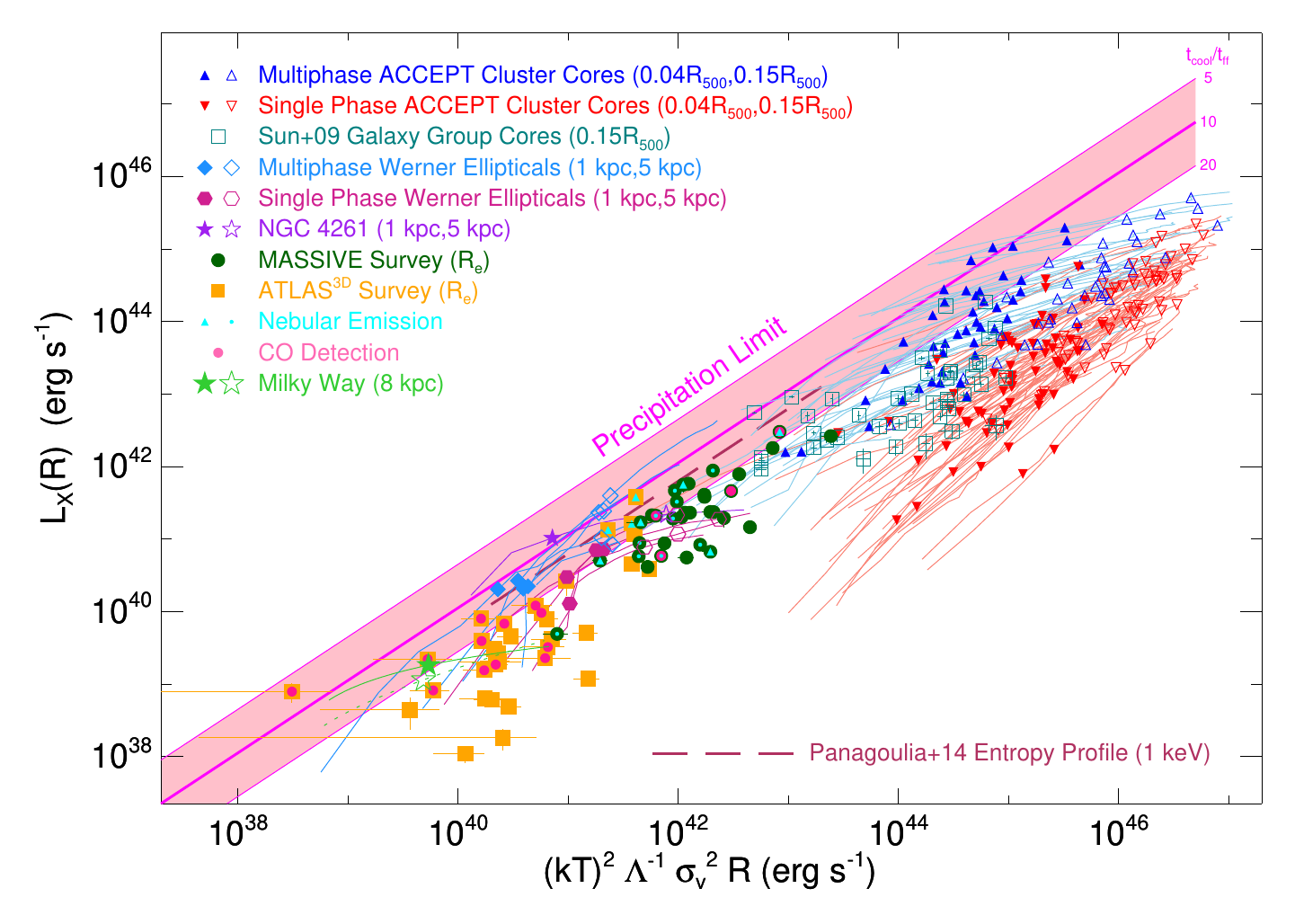} \\
\end{center}
\caption{ \footnotesize 
Comparison of the X-ray luminosity $L_X(R)$ from within radius $R$ to the precipitation-limited luminosity $(9 \pi / 25) (kT)^2 \Lambda^{-1} \sigma_e^2 R$ (thick magenta line) obtained from the condition $\min ( t_{\rm cool} / t_{\rm ff} ) \gtrsim 10$.  The pink region bounded by thinner lines shows the range corresponding to $5 < \min ( t_{\rm cool} / t_{\rm ff} ) < 20$.  Other lines show tracks of $L_X(R)$ that progress from lower left to upper right as $R$ increases.  Blue lines with triangles represent ACCEPT cluster cores with H$\alpha$ emission, and red lines with triangles represent ACCEPT cluster cores with optical spectroscopy but no detectable H$\alpha$.  For the ACCEPT clusters, solid triangles indicate $R = 0.04 R_{500}$ and open triangles indicate $0.15 R_{500}$.  %Bold lines show model tracks characteristic of galaxy clusters with core entropy corresponding to $10 \, {\rm keV \, cm^2}$ (dot-dashed gray line), $30 \, {\rm keV \, cm^2}$ (solid gray line), and $100 \, {\rm keV \, cm^2}$ (dashed gray line).  
Teal squares show the luminosities of group cores (within $0.15 R_{500}$) from \citet{Sun+09}.  Lines with diamonds, hexagons, and stars represent tracks based on observations by \citet{Werner+2012MNRAS.425.2731W,Werner+2014MNRAS.439.2291W} of massive ellipticals with extended H$\alpha$ emission (blue diamonds) and without extended H$\alpha$ emission (red hexagons).  Purple stars indicate the massive elliptical NGC~4261, which hosts a particularly powerful AGN outflow.  For these massive ellipticals, solid symbols indicate $R = 1$~kpc and open symbols indicate $R = 5$~kpc.  A dashed line shows the locus derived from the \citet{Panagoulia_2014MNRAS.438.2341P} entropy profile from 0.1~kpc to 100~kpc for $kT = 1$~keV. The Milky Way tracks in green are based on the fits of Miller \& Bregman (2015) to O~VIII emission observations (solid line and solid star) and O~VII emission observations (dotted line and open star), with stars indicating $R = 8$~kpc.  Solid points without tracks are from Goulding et al. (2016) and represent $L_X$ measured within an aperture corresponding to the effective radius for early-type galaxies from the MASSIVE (green circles) and ATLAS$^{\rm 3D}$ (orange squares) surveys.  Those measurements may exceed the actual $L_X(R)$ values by up to 50\% because of projected flux from regions at $> R_e$.  Symbols overplotted on the MASSIVE and ATLAS$^{\rm 3D}$ points indicate galaxies showing evidence for multiphase gas, with small cyan triangles and dots indicating detections of nebular line emission and larger pink circles indicating CO detections.
\vspace*{1em}
\label{fig-LX_Lprecip}}
\end{figure*}
% ----------------------------------------

Our comparison in Figure~\ref{fig-LX_Lprecip} of observations with the limiting luminosity predicted by equation (\ref{eq-LXTRsigv}) draws on several different data sets:
\begin{itemize}

\item {\bf Galaxy-cluster cores.} The ACCEPT database \citep{Cavagnolo+09} is our source for the luminosities and temperatures of galaxy-cluster cores.  We use the subset analyzed by Voit et al. (2015), which separates into two categories:  (1) multiphase clusters with detectable H$\alpha$ emission (blue lines with triangles), and (2) single-phase clusters with optical spectroscopy showing no detectable H$\alpha$ emission (red lines with triangles).  In order to compute the precipitation limit on $L_X(R)$, we use an approximation to $\sigma_v(R)$ consisting of the sum of a singular isothermal sphere with $\sigma_v = 300 \, {\rm km \, s^{-1}}$ and an NFW profile \citep{nfw97} of fixed concentration and an amplitude determined by the cluster's X-ray temperature.  Concentration is defined with respect to the radius $R_{500}$ that encloses a mean matter density 500 times the critical density, and the scale radius at which the mass-density profile is $\propto R^{-2}$ is assumed to be $R_s = R_{500} / 3$.  We show tracks of $L_X(R)$ that go from $R \leq 10$~kpc through $R = 300$~kpc, but focus particularly on measurements of $L_X(R)$ within the apertures $0.04 R_{500}$ (filled triangles) and $0.15 R_{500}$ (open triangles).  The larger aperture is commonly used to separate the core regions of a cluster from its outer parts.  The smaller aperture yields high signal-to-noise measurements of core properties.  For a 5~keV galaxy cluster, our model for the gravitational potential gives $\sigma_v(0.04 R_{500}) = 454 \, {\rm km \, s^{-1}}$, $\sigma_v(0.15 R_{500}) = 630 \, {\rm km \, s^{-1}}$, and $\max ( \sigma_v ) = 760 \, {\rm km \, s^{-1}}$.

\item {\bf Galaxy-group cores.} Our primary source for the luminosities and temperatures of galaxy-group cores (teal squares) is \citet{Sun+09}.  For these objects, we use the luminosity within the radius $0.15 R_{500}$, and $R_{500}$ is obtained from a hydrostatic mass model for each group.  We do not have direct measurements of $\sigma_v(0.15 R_{500})$ and so instead use equation (\ref{eq-LXTRbeta}) with $\beta = 0.5$ to obtain precipitation limits on $L_X(R)$. Our secondary source is the \citet{Panagoulia_2014MNRAS.438.2341P} sample of groups and clusters.  That compilation contains 13 systems with $kT \approx 1$~keV that have a mean entropy profile $K(R) = 95.4 \, {\rm keV \, cm^2} \times (R/{\rm 100 \, kpc})^{2/3}$, where $K \equiv kTn_e^{-2/3}$.  Representative values of $L_X(R)$ and precipitation limits can be derived for these systems by setting $kT = 1$~keV and applying equation (\ref{eq-LXTRbeta}) with $\beta = 0.5$.

\item {\bf Werner ellipticals.}  High-quality {\em Chandra} observations of ten nearby massive elliptical galaxies were presented by \citet{Werner+2012MNRAS.425.2731W,Werner+2014MNRAS.439.2291W} and analyzed by \citet{Voit+2015ApJ...803L..21V} in the context of precipitation-regulated feedback.  Five of them (blue lines with diamonds) contain multiphase gas that extends over several kpc.  The other five (mostly red lines with hexagons) are considered single-phase galaxies because there is no observable multiphase gas beyond the central kpc.  However, one of them (NGC 4261, purple line with stars) contains a central sub-kpc disk of cool, dusty gas that presumably fuels the galaxy's strong bipolar outflow, which is roughly two orders of magnitude more powerful than the AGN outflows in the other nine galaxies.  Deprojected density and temperature profiles from those studies are used to reconstruct $L_X(R)$. Precipitation limits on $L_X(R)$ are computed for those galaxies using $\sigma_v$ values from the Hyperleda database\footnote{http://leda.univ-lyon1.fr/} under the assumption that $\sigma_v(R)$ is constant.  The lines go from $R < 1$~kpc to $R > 10$~kpc, with endpoints limited by data quality.  Filled symbols mark $R = 1$~kpc and open ones indicate $R = 5$~kpc. 

\item {\bf Volume-limited Surveys.}  Ambient gas in early-type galaxies of lower mass is harder to observe because of its lower surface brightness.   The radii out to which diffuse X-ray emission can be reliably measured are correspondingly smaller, and {\em Chandra} observations are required in order to excise the X-ray emission from individual X-ray binaries.  Here, we take advantage of a {\em Chandra} archival analysis by \citet{Goulding_2016ApJ...826..167G} of 74 early-type galaxies from two volume-limited samples: ATLAS$^{\rm 3D}$ ($M_* > 10^{9.9} M\odot$ and within $42$~Mpc) and MASSIVE ($M_* > 10^{11.5} M\odot$ and within 108~Mpc).  Measurements of each galaxy's bolometric\footnote{PIMMS was used to calculate a bolometric correction to the 0.3-5~keV $L_X$ measurements of \citet{Goulding_2016ApJ...826..167G}.} X-ray luminosity within an aperture corresponding to the effective radius ($R_e$) of starlight are shown with orange squares (ATLAS$^{\rm 3D}$) and green circles (MASSIVE).  Some of those galaxies also contain detectable multiphase gas.  Overplotted pink dots indicate galaxies with CO detections from \citet{Davis_2011MNRAS.417..882D,Davis_2016MNRAS.455..214D}.  Cyan symbols indicate galaxies with detectable nebular optical line emission \citep{Pandya_2017ApJ...837...40P}.  Among the cyan symbols, triangles represent extended emission, while circles represent unresolved detections.  

\item{\bf Milky Way.}  Our $L_X(R)$ tracks for the Milky Way (green lines with stars) are derived from the models of Miller \& Bregman (2015), which represent fits to {\em Chandra} observations of O~VII (dotted line and open star) and O~VIII (solid line and filled star) emission lines along many different lines of sight through the Galaxy.  We use the best fitting parameters in their optically-thin models.  The tracks go from $R = 1$~kpc to 10~kpc, and stars indicate $R = 8$~kpc.

\end{itemize}

\subsection{The Precipitation Limit}

The main finding of this paper is that the precipitation limit derived from assuming $\min (t_{\rm cool} / t_{\rm ff}) = 10$ tracks the upper envelope of the combined data set over nearly seven orders of magnitude in $L_X$.  Figure~\ref{fig-LX_Lprecip} presents all of the data on a single plot, along with a magenta line showing the precipitation limit $(9\pi/25) (kT)^2 \Lambda^{-1} \sigma_v^2 R$ for solar-metallicity gas.  The pink area shows a factor of 4 range around that limit, corresponding to $5 < \min (t_{\rm cool} / t_{\rm ff}) < 20$.  The brightest ACCEPT clusters and Werner ellipticals all come quite close to the precipitation limit at small radii, as we have already shown elsewhere (Voit et al. 2015a,b).  However, the fact that the upper envelopes of both the ATLAS$^{\rm 3D}$ and MASSIVE samples also track the precipitation limit, along with the galaxy-group cores from \citet{Sun+09}, is a new result.  Apparently, the physical processes responsible for regulating galaxy cluster cores may also be regulating the hot ambient media of smaller galaxies, all the way down through Milky Way scales. 

Another notable feature of Figure~\ref{fig-LX_Lprecip} is that no break is evident in the relation between $L_X(R)$ and $(kT)^2 \Lambda^{-1} \sigma_v^2 R$, even though $\Lambda \propto T^{1/2}$ at the high end of the temperature range and $\Lambda \propto T^{-1}$ at the low end.  The absence of a break indicates that the upper limit on $L_X$ at a given $T$ is set primarily by radiative cooling, because the presence of $\Lambda (T)$ in the denominator of the precipitation limit on $L_X(R)$ compensates for the increasing steepness generally seen in the $L_X$--$T$ relation at fixed $R$ as one moves from clusters through groups down to individual early-type galaxies.   However, much of the X-ray luminosity data available in the literature for early-type galaxies does not specify a fixed metric aperture.  

% ----------------------------------------
\begin{figure*}[th]
\begin{center}
\includegraphics[width=7in, trim = 0.1in 0.1in 0.0in 0.0in]{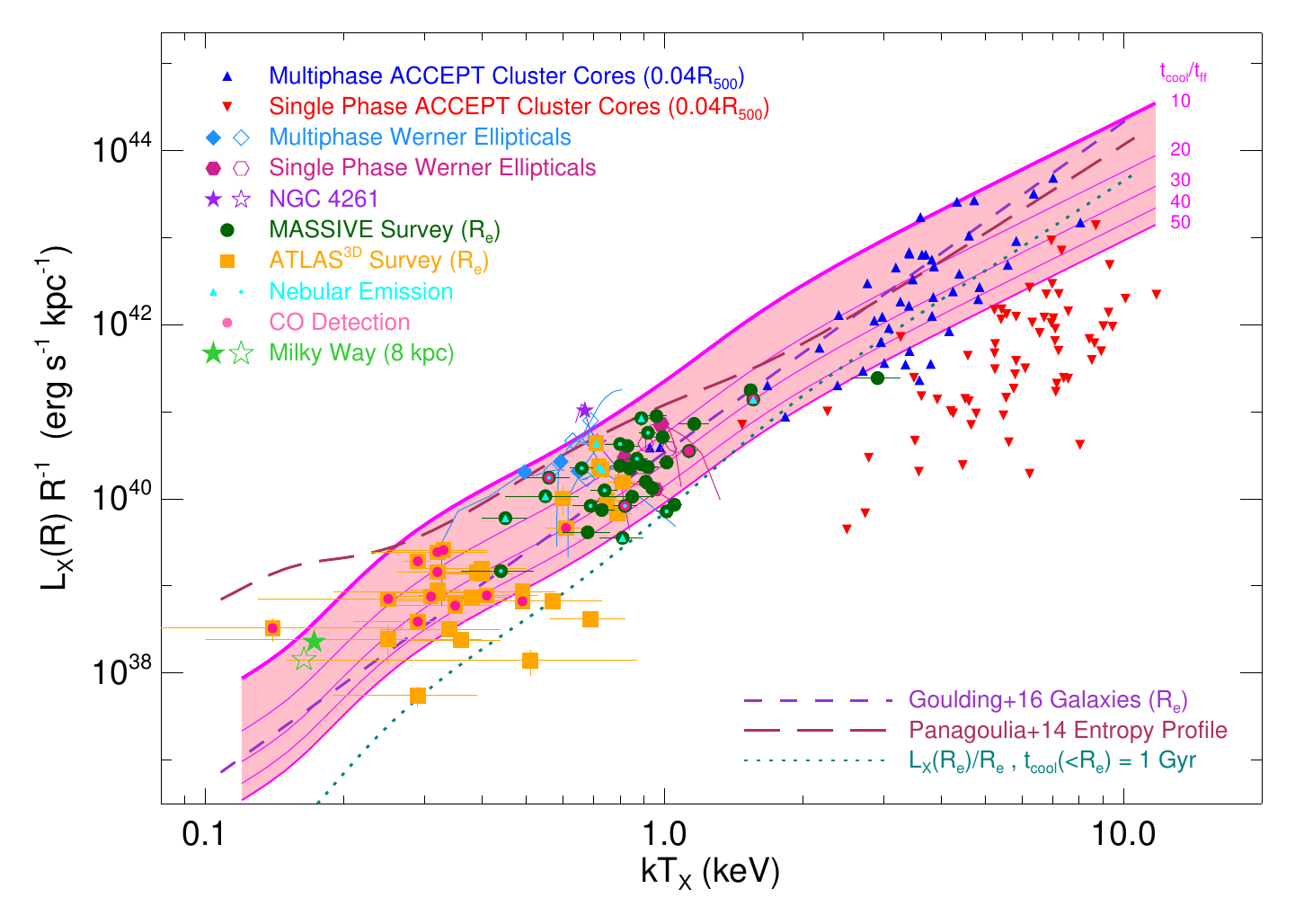} \\
\end{center}
\caption{ \footnotesize 
Relationship between $L_X(R) \cdot R^{-1}$ and $T$ within the stellar envelopes of early-type galaxies.  Symbols represent the same data as in Figure~\ref{fig-LX_Lprecip}.  The short-dashed line shows the best $L_X$--$T$ fit from Goulding et al. (2016) to the MASSIVE and ATLAS$^{\rm 3D}$ samples, divided by $R_e = (8.8 \, {\rm kpc}) (kT/1 \, {\rm keV})^{0.9}$, which is a fit to the $R_e$--$T$ relation for those same galaxies.   Solid magenta lines show tracks of $L_X(R) \cdot R^{-1}$ as functions of $kT$ given by equation (\ref{eq-LX_invR_T}) for different values of $t_{\rm cool}/t_{\rm ff}$, as labeled.  The long-dashed line shows a track derived from the entropy profile measured by  \citet{Panagoulia_2014MNRAS.438.2341P} among early-type galaxies.  A dotted line shows the $L_X(R_e) \cdot R_e^{-1}$ track for systems with a constant cooling time $t_{\rm cool} = 1$~Gyr within $R_e = (8.8 \, {\rm kpc}) (kT/1 \, {\rm keV})^{0.9}$.  Notice that the trend among multiphase systems is to follow the magenta lines, which have power-law slope $\sim T^4$ for $kT \lesssim 1$~keV but gradually flatten to $\sim T^{2.5}$ at greater temperatures.  While multiphase galaxies and galaxy-cluster cores tend to remain within the pink region ranging from $t_{\rm cool}/t_{\rm ff} = 10$ to 50, single-phase systems can have lower X-ray luminosities.
\vspace*{1em}
\label{fig-LX_invR_T}}
\end{figure*}
% ----------------------------------------

One must therefore account for the dependence of aperture radius $R$ on $T$, and the limiting $L_X$--$T$ relation changes accordingly.   Specifying an aperture with a radius equal to the effective radius $R_e$ of the galaxy's starlight allows a well-defined $R_e(T)$ relation to be derived from observations of systems with $kT \lesssim 1$~keV.  For the \citet{Goulding_2016ApJ...826..167G} collection of ATLAS$^{\rm 3D}$ and MASSIVE galaxies, the approximate relation is $R_e \approx (8.8 \, {\rm kpc}) (kT/1 \, {\keV})^{0.9}$.  The precipitation-limited luminosity from within $R_e$ therefore scales as $L_X(R_e) \propto T^{\zeta}$, with $\zeta \approx 4.4$--4.9 in the temperature range $0.2 \, {\rm keV} \lesssim kT \lesssim 1.0 \, {\rm keV}$, within which the power-law scaling of $\Lambda \propto T^\lambda$ is $\lambda \approx -1$ to $-0.5$.  In their best fit to the observed $L_X(R_e)$--$T$ relation in this temperature range, \citet{Goulding_2016ApJ...826..167G} found $\zeta = 4.7$.  \citet{KimFabbiano_2015ApJ...812..127K} found $\zeta = 4.4$--4.6 for early-type galaxies with cored stellar light profiles in a similar archival {\em Chandra} study.  

\subsection{The $L_X(R_e)$--$T$ Relation}

Figure 2 shows the luminosity-temperature relation that results when $L_X(R)$ is divided by the aperture radius $R$ in order to remove the aperture dependence of the precipitation limit.  Most of the galaxy points correspond to the aperture $R_e$.  The cluster-core points correspond to $0.04 R_{500}$, which is typically $\sim 40$~kpc.\footnote{We are using $0.04 R_{500}$ as a proxy for $R_e$ because the stellar envelopes of central cluster galaxies are so extended that their effective radii are not yet well defined \citep[e.g.,][]{Kravtsov_2014_BCG_arXiv1401.7329K}.}   A short-dashed purple line shows the $L_X(R_e)$--$T$ fit from \citet{Goulding_2016ApJ...826..167G}, divided by the $R_e(T)$ fit from the previous paragraph.  Magenta lines show tracks derived from a version of equation (\ref{eq-LXTRbeta}) that has been generalized to make $t_{\rm cool} / t_{\rm ff}$ a free parameter and in which $\beta = 0.5$ to accord with the power-law slope of the precipitation-limited pressure profile:
\begin{equation}
  \frac {L_X(R)} {R} = \frac {18 \pi} {\mu m_p} \frac {(kT)^3} {\Lambda(T)} 
  	\left( \frac {t_{\rm cool}} {t_{\rm ff}} \right)^{-2}
	\; \; .
  \label{eq-LX_invR_T}
\end{equation}
Making this generalization allows the figure to show how the data points reflect the typical value of $t_{\rm cool} / t_{\rm ff}$ corresponding to a particular combination of $L_X(R) \cdot R^{-1}$ and $kT$. For comparison, a teal dotted line shows
\begin{equation}
  \frac {L_X(R_e)} {R_e} = \frac {12 \pi R_e^2} {({\rm 1 \, Gyr})^2} 
  					\frac {(kT)^2} {\Lambda(T)}
	\; \; ,
  \label{eq-LX_invR_T}
\end{equation}
for $R_e \approx (8.8 \, {\rm kpc}) (kT/1 \, {\keV})^{0.9}$, which describes systems with a constant cooling time $t_{\rm cool} = 1$~Gyr inside of $R_e$.

The track corresponding to $t_{\rm cool}/t_{\rm ff} = 10$ again follows the upper envelope of the data points but now changes in slope because of the $\Lambda(T)$ factor in equation (\ref{eq-LX_invR_T}).  Most of the data points follow the same trend with $T$ but correspond to greater $t_{\rm cool}/t_{\rm ff}$ values, centered near $t_{\rm cool}/t_{\rm ff} \approx 25$.  This finding supports the hypothesis that the change in $L_X$--$T$ slope observed near $\sim 1$~keV among early-type galaxies is caused by precipitation-regulated feedback.  While ratios as great as $t_{\rm cool}/t_{\rm ff} \gtrsim 25$ may seem too large for condensation and precipitation, we would like to emphasize that the MASSIVE and ATLAS$^{\rm 3D}$ points indicate {\em average} values of $t_{\rm cool}/t_{\rm ff}$ within $R_e$.  The minimum values of $t_{\rm cool}/t_{\rm ff}$ may be considerably smaller, especially near the center.  For examples, see the tracks belonging to the Werner ellipticals in Figure~\ref{fig-LX_Lprecip}, which are close to the precipitation limit at $\sim 1$~kpc but can diverge from it at larger radii.

Both \citet{KimFabbiano_2015ApJ...812..127K} and \citet{Goulding_2016ApJ...826..167G} point out that recent simulations by \citet{Choi_2015MNRAS.449.4105C} obtain a similar $L_X$--$T$ relation for early-type galaxies with $kT \lesssim 1$~keV when the AGN feedback is kinetic instead of thermal.  This change in feedback mode reduces $L_X$ by two orders of magnitude in their simulations, but why does that happen?  \citet{Meece_2017ApJ...841..133M} have shown that changing the mode of AGN energy injection from thermal to kinetic has deep implications for precipitation and self-regulation.  Pure thermal energy injection at a galaxy's center fails to bring about self regulated AGN feedback because it overturns the surrounding entropy gradient.  The resulting convection promotes thermal instability and triggers runaway condensation.
 
\citet{Voit_2017_BigPaper} discuss these issues in detail and show that self-regulation requires an energy-injection mechanism that deposits heat into the circumgalactic medium without inverting its entropy gradient.  Bipolar AGN jets are one such energy injection mechanism, but the key feature is not the kinetic energy.  Instead, it is the response of the global entropy gradient to energy input.  If feedback maintains a rising entropy profile outside the central few kpc, then self-regulation through precipitation will naturally produce the observed $L_X$--$T$ relation.

Models that assume a universal power-law entropy profile are less successful.  For example, \citet{Panagoulia_2014MNRAS.438.2341P} find that the mean profile $K(R) = 95.4 \, {\rm keV \, cm^2} \times (R/{\rm 100 \, kpc})^{2/3}$ derived from systems with $kT \approx 1$~keV is also an adequate description of many galaxy-cluster cores.  However, the relation derived from a universal entropy profile with this slope scales as $L_X(R) \propto T^3 \Lambda(T) R$, as shown by the long-dashed line in Figure \ref{fig-LX_invR_T}.  Its slope is similar to the precipitation limit in the $kT \approx 0.3$--1.0 keV range but diverges from the precipitation slope at both higher and lower temperatures.  Figure \ref{fig-LX_invR_T} also demonstrates that the \citet{Panagoulia_2014MNRAS.438.2341P} profile cannot be universal, because of the large dispersion in $L_X(R) \cdot R^{-1}$ at each temperature.  The cluster-core points in particular span more than three orders of magnitude in $L_X(R) \cdot R^{-1}$, which requires the core entropy of the ambient medium to span at least an order-of-magnitude range in objects of a given temperature. 

\subsection{Incidence of Multiphase Gas}

Another feature of Figures \ref{fig-LX_Lprecip} and \ref{fig-LX_invR_T} indicates a potentially interesting avenue for further exploration:  The presence of multiphase gas in MASSIVE and ATLAS$^{\rm 3D}$ galaxies correlates with their proximity to the precipitation limit.  All of those multiphase galaxies are within or on the margins of the pink region in Figure \ref{fig-LX_invR_T}.  Likewise, the ACCEPT points representing multiphase galaxy-cluster cores occupy the same region, while the single-phase cluster cores generally lie below it, as do several of the single-phase ATLAS$^{\rm 3D}$ galaxies. 

We have shown elsewhere that the radially resolved $t_{\rm cool}(R)$ profiles of galaxy clusters are strongly dichotomous, with $10 \lesssim \min (t_{\rm cool}/t_{\rm ff}) \lesssim 20$ typical among the multiphase population and $\min (t_{\rm cool}/t_{\rm ff}) \gtrsim 20$ among the single-phase population, with few exceptions \citep{VoitDonahue2015ApJ...799L...1V,Voit_2015Natur.519..203V}.  \citet{Hogan_2017_tctff} have corroborated this finding among galaxy-cluster cores, while \citet{Voit+2015ApJ...803L..21V} have presented evidence that it also holds among massive ellipticals with $kT \approx 1$~keV.  Figures \ref{fig-LX_Lprecip} and \ref{fig-LX_invR_T} suggest that this trend may extend to systems of even lower temperature, but spatially-resolved $t_{\rm cool}(R)$ profiles will be needed to verify that conjecture.

\section{A General $L_X$--$T$--$R$ Model out to $R_{500}$}
\label{sec-LX_T_R500}

\begin{figure*}[t]
\begin{center}
\includegraphics[width=7in, trim = 0.1in 0.1in 0.0in 0.0in]{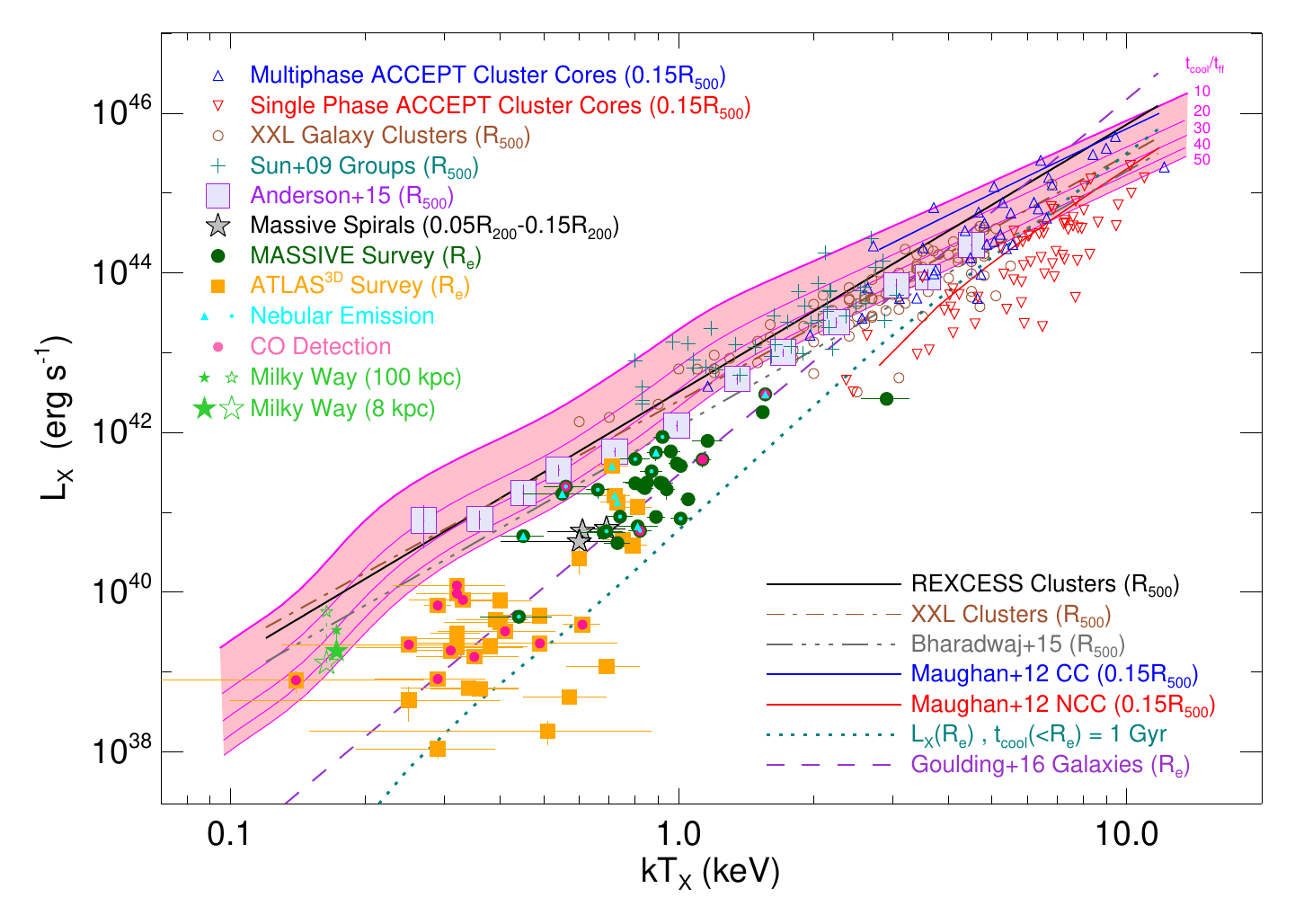} \\
\end{center}
\caption{ \footnotesize 
Comparison of the $L_X$--$T$ relation within $R_{500}$ to the $L_X$--$T$ relations at smaller radii. Many of the symbols represent the same data as in Figure~\ref{fig-LX_Lprecip} but some are new.  Teal crosses represent group luminosities from \citep{Sun+09} within $R_{500}$ (instead of $0.15 R_{500})$.  Brown circles represent $L_X(R_{500})$ for clusters and groups from the XXL survey \citep{Giles_2016_XXL_LX-T}, which have been corrected for redshift evolution.  Squares with purple borders represent the $L_X(R_{500})$--$T$ relation derived from stacked ROSAT observations by \citet{Anderson_2015MNRAS.449.3806A}.  Stars with black borders show $L_X$ and $T$ measured by \citet{Bogdan_2013ApJ...772...97B,Bogdan_2013ApJ...772...98B} within the annulus (0.05--0.15)$R_{200}$ around three massive spiral galaxies.  A solid black line shows the ``orthogonal" $L_X(R_{500})$--$T$ fit to the REXCESS clusters from \citet{Pratt_2009_REXCESS_LX-T}, which is $\propto T^{3.35}$.   A dot-dashed brown line shows the best fitting bias-corrected relation for the XXL clusters ($\propto T^{3.08}$).  A gray dot-dot-dot-dashed line shows a bias-corrected relation ($\propto T^{3.2}$) from \citet{Bharadwaj_2015A&A...573A..75B}.  The other lines correspond to smaller apertures.  Blue and red lines show $L_X(0.15R_{500})$--$T$ relations derived from the fits of \citep{Maughan_2012_LX-T} to cool-core clusters and non-cool-core clusters, respectively.  A dashed violet line shows the best-fit $L_X(R_e)$--$T$ relation from \citep{Goulding_2016ApJ...826..167G}.  A dotted teal line shows $L_X$ from within $R_e = (8.8 \, {\rm kpc}) (kT/{\rm 1 \, kpc})^{0.9}$ for gas with a constant cooling time $t_{\rm cool} = 1$~Gyr. Magenta lines and the pink region show predicted $L_X(R_{500})$--$T$ relations derived from the simple model of \S \ref{sec-LX_T_R500} for different limiting values of $t_{\rm cool}/t_{\rm ff}$, as labeled.
\vspace*{1em}
\label{fig-LX_T}}
\end{figure*}

Our objective in this section is to present a maximally simple model for the overall $L_X$--$T$ relation that incorporates the precipitation limit \citep[see also][]{Sharma+2012MNRAS.427.1219S}.  The standard aperture for this relation among groups and clusters of galaxies is $R_{500}$.  X-ray surface brightness profiles of massive galaxy clusters are frequently well-observed out to at least this radius \citep[e.g.,][]{Maughan_2012_LX-T}.  Galaxy groups can also be observed out to this radius but require a careful treatment of the soft X-ray backgrounds \citep[e.g.,][]{Sun+09}.  X-ray emission from individual galaxies cannot be measured out to $R_{500}$, but a mean $L_X(R_{500})$--$T$ relation for massive galaxies can be obtained by stacking ROSAT observations of large numbers of galaxies \citep{Anderson_2015MNRAS.449.3806A}.\footnote{The \citet{Anderson_2015MNRAS.449.3806A} $L_X$--$T$ points presented in this paper differ from the original ones in two ways: (1) we perform our own bolometric correction using PIMMS, and (2) we adjust the original temperature scale downward by 10\% to account for the halo-mass recalibration by \citet{Wang_2016MNRAS.456.2301W}.}  Figure \ref{fig-LX_T} shows a representative selection of such measurements.  For comparison, the figure also shows the $L_X$--$T$ points from Figures~\ref{fig-LX_Lprecip} and \ref{fig-LX_invR_T}, which represent measurements within smaller apertures.  Those points show that the slope of the observed $L_X$--$T$ relation at $\lesssim 2$~keV is aperture dependent.  Within the aperture $R_{500}$, the slope of the $L_X$--$T$ relation remains nearly constant over the entire observed temperature range.
 
% ----------------------------------------
% ----------------------------------------

Pink shading in Figure \ref{fig-LX_T} shows a model for the $L_X(R_{500})$--$T$ relation that combines a cosmological entropy profile at large radii with a precipitation-limited entropy profile at small radii.  This model accounts for the lack of a break in slope.  The cosmological entropy profile is extremely simple.  It assumes that the gravitational potential is a singular isothermal sphere in which the baryon mass fraction at each radius is equal to the cosmological baryon fraction.  In hydrostatic equilibrium, the gas temperature is $kT = \mu m_p \sigma_v^2$, giving an entropy profile
\begin{equation}
  K_{\rm cos}(R) = \mu m_p \sigma_v^2 
  				\left[ \frac {125 f_b H^2(z)} {2 \pi G \mu_e m_p} \right]^{-2/3} 
				\left( \frac {R} {R_{500}} \right)^{4/3} \; ,
\end{equation}
%\begin{eqnarray}
%  K_{\rm cos}(r) & = & \mu m_p \sigma_v^2 
%  				\left[ \frac {125 f_b H^2(z)} {2 \pi G \mu_e m_p} \right]^{-2/3} 
%				\left( \frac {R} {R_{500}} \right)^{4/3} \; \; , \\
%			& = & (220 \, {\rm keV \, cm^2}) 
%				\left( \frac {\sigma_v} {300 \, {\rm km \, s^{-1}}} \right)^2
%				\left( \frac {f_b} {0.16} \right)^{-2/3}
%				E^{-4/3}(z) \left( \frac {R} {R_{500}} \right)^{4/3} \; \; .
%\end{eqnarray}
where $H(z)$ is the Hubble expansion parameter and $\mu_e$ is the mean mass per electron.  Without modification, gas with this entropy profile would produce an X-ray luminosity that diverges at small radii.  In non-radiative cosmological simulations, mixing produces an entropy core that keeps the X-ray luminosity from diverging \citep[e.g.,][]{Mitchell+09}, but when radiative cooling is turned on, $L_X$ is limited by a combination of cooling and feedback \citep[e.g.,][]{Nagai+2007ApJ...655...98N}.

In the precipitation framework, the limiting entropy profile at small radii depends on the $t_{\rm cool}/t_{\rm ff}$ ratio at which condensation-triggered feedback self regulates.  Consequently, the X-ray luminosity of a precipitation regulated system is determined by the limiting value of $t_{\rm cool}/t_{\rm ff}$.  The electron density profile corresponding to this limit is given by a generalization of equation (\ref{eq-max_ne}): 
\begin{equation}
  n_{e,{\rm pre}}(R) \: = \: \frac {3 kT \sigma_v} {\Lambda(T)} \left( \frac {n} {2 n_i} \right) 
             \left( \frac {t_{\rm cool}} {t_{\rm ff}} \right)^{-1} R^{-1}
    \; \; .
    \label{eq-ne}
\end{equation}
The temperature of hydrostatic, isothermal gas with this density profile is $kT = 2 \mu m_p \sigma_v^2$ in an isothermal potential and gives the limiting entropy profile
\begin{equation}
  K_{\rm pre}(R) = (2 \mu m_p)^{1/3} \left[ \frac {2 n_i \Lambda(T)} {3 n} \right]^{2/3} 
		                \left( \frac {t_{\rm cool}} {t_{\rm ff}} \right)^{2/3} 
				R^{2/3} \; \; . 
				\vspace*{0.1em}
\end{equation}
%\begin{eqnarray}
% K_{\rm precip}(r) & = & 3^{2/3} (2 \mu m_p \sigma_v^2)^{1/3} \Lambda(T_{\rm precip}) 
%		                \left( \frac {t_{\rm cool}} {t_{\rm ff}} \right)^{2/3} 
%				\left( \frac {R} {\sigma_v} \right)^{2/3} \; \; , \\
%			& = & (91 \, {\rm keV \, cm^2}) 
%				\left( \frac {\sigma_v} {300 \, {\rm km \, s^{-1}}} \right)^{-2/3}
%				\left( \frac {\Lambda(T_{\rm precip})} 
%					{10^{-23} \, {\rm erg \, cm^3 \, s^{-1}} \right)^{2/3}
%		                \left( \frac {t_{\rm cool} / t_{\rm ff}} {10} \right)^{2/3} 
%				E^{-2/3}(z) \left( \frac {R} {R_{500}} \right)^{2/3} \; \; .
%\end{eqnarray}
When calculating $K_{\rm pre}(r)$, we use a cooling function with solar metallicity, because AGN triggering in a massive galaxy happens in regions where the gas-phase abundances are approximately solar.  The X-ray luminosity of gas with this entropy profile diverges toward large radii and is ultimately limited by the cosmological entropy profile.

Adding the generalized precipitation profile to the cosmological profile gives a combined entropy profile that depends only on $\sigma_v$ and the limiting value of $t_{\rm cool}/t_{\rm ff}$:
\begin{equation}
  K(R) = K_{\rm cos}(R) + K_{\rm pre}(R)  \; \; .
\end{equation}
The power-law slope of this combined profile is $K \propto R^{2/3}$ at small radii and asymptotically approaches $K \propto R^{4/3}$ at large radii.  In hydrostatic equilibrium in an isothermal potential well, the pressure profile of a nearly isothermal gas is $d \ln P / d \ln r \approx - (3/2) (d \ln K / d \ln R)$.  One can therefore obtain an approximate temperature profile
\begin{equation}
  kT(R) =  \frac {\mu m_p \sigma_v^2 \; K(R)} {K_{\rm cos}(R) + 0.5 K_{\rm pre}(R)}
\end{equation}
from the logarithmic derivative of $K(R)$, which leads to the approximate electron density profile \begin{equation}
  n_e(R) = \left[ \frac {K_{\rm cos}(R) + 0.5 K_{\rm pre}(R)} {\mu m_p \sigma_v^2} \right]^{-3/2}
  \; \; .
\end{equation}  
Integrating equation (\ref{sec-LX(R)}) with this density profile then gives the model value of $L_X(R_{500})$.  The model value of $T$ is simply the emissivity-weighted average of $T(R)$.  In both of these integrations, the cooling function $\Lambda(T)$ depends strongly on metallicity for $kT \lesssim 2$~keV.  We therefore assume a representative metallicity profile $Z(R)/Z_\odot = \min[1.0,0.3 (R/R_{500})^{-1/2}]$ that is broadly consistent with measurements of galaxy groups \citep[e.g.,][]{RasmussenPonman_2007MNRAS.380.1554R,Sun+09}.

Magenta lines in Figure~\ref{fig-LX_T} are $L_X(R_{500})$--$T$ tracks corresponding to different values of $t_{\rm cool} /t_{\rm ff}$ in equation (\ref{eq-ne}).  This ratio is the only free parameter in the model and specifies the asymptotic value of $t_{\rm cool} /t_{\rm ff}$ at small radii.  Setting $t_{\rm cool} /t_{\rm ff} = 10$ gives a track that follows the upper envelope of the observations from $\approx 10$~keV down through $\lesssim 1$~keV, at which X-ray measurements extending out to $R_{500}$ in individual objects become extremely difficult.  That track changes slope in two places.  A subtle change happens near 1~keV.  The mean value of $\beta = \mu m_p \sigma_v^2 / kT$ rises from $\approx 0.5$ to $\approx 1$ as $kT$ increases through $\sim 1$~keV.  This change in $\beta$ shifts points with $kT \lesssim 1$~keV to greater temperatures for a given $\sigma_v$, which causes the $L_X(R_{500})$--$T$ relation to steepen near 1~keV.   A more pronounced change in slope happens near 0.2~keV, below which the entire profile out to $R_{500}$ is essentially the same as the precipitation-limited profile.  In that regime, which includes the Milky Way, the model predicts $L_X(R_{500}) \propto T^{3.5} \Lambda^{-1}(T)$, with $\Lambda \appropto T^{-1}$. 

Each of the data sets in Figure~\ref{fig-LX_T} suffers from biases that we will not fully analyze here.  For example, surveys of X-ray selected clusters and groups of galaxies tend to be biased toward the more X-ray luminous examples in each mass and temperature range and overestimate the mean $L_X(R_{500})$ at a given $T$ if that bias is not properly corrected \citep[e.g.,][]{Allen_2011ARA&A..49..409A}.  Conversely, optically-selected samples can tend to underestimate $L_X(R_{500})$ at a given $T$ because of how scatter in the relation between the optical mass proxy and the actual system mass combines with the steep slope of the mass function \citep[e.g.,][]{Rozo_2014MNRAS.438...78R}.  A blend of these two biases is responsible between the offset between the $L_X(R_{500})$--$T$ relations derived from the X-ray selected samples (REXCESS and XXL) and the $L_X(R_{500})$--$T$ relation derived by \citet{Anderson_2015MNRAS.449.3806A} from ROSAT stacks of luminous red galaxies from the Sloan Digital Sky Survey.  An additional contribution to the offset may come from anticorrelation of $L_X$ with galactic stellar mass at fixed total mass.

However, despite those biases, the slopes of the $L_X(R_{500})$--$T$ relations derived from REXCESS and XXL remain similar to the slope of the \citet{Anderson_2015MNRAS.449.3806A} points down to temperatures at least a factor of 2 lower the the bottom end of the range covered by the X-ray surveys, and maybe even down to the Milky Way itself.  Furthermore, the $L_X(R_{500})$--$T$ relation derived from our simple model shares the same slope.  Taken together, these results strongly suggest that feedback regulation of circumgalactic gas around galactic systems with $kT \approx 0.2$--1.0~keV shares much in common with the AGN feedback mechanism that regulates the cores of massive galaxy clusters.

In galaxy clusters, strong AGN feedback is tightly linked to the presence of a cool core, which in turn is linked with the presence of multiphase gas in the central galaxy \citep{mn07,Cavagnolo+08,McNamaraNulsen2012NJPh...14e5023M}.  \citet{Maughan_2012_LX-T} have measured separate $L_X(R_{500})$--$T$ relations for cool-core and non-cool-core clusters, as well as $L_X(R_{500})$--$T$ relations excluding the core region within $0.15 R_{500}$.  When the cores are excluded, the $L_X(R_{500})$--$T$ relations for cool-core ($\propto T^{2.1}$) and non-cool-core clusters ($\propto T^{2.9}$) overlap one another, with a slope $\propto T^{2.7}$ for the combined sample.  These populations overlap because the radial profiles of entropy and electron density beyond $0.15 R_{500}$ are nearly identical to the cosmological profile for clusters with $kT \gtrsim 3.5$~keV.  However, the core regions of those two cluster populations are quite different.

Figure~\ref{fig-LX_T} shows the $L_X(0.15 R_{500})$--$T$ relations we obtain from \citet{Maughan_2012_LX-T} by subtracting the core-excluded relations from the full-aperture relations.  The lines for cool-core and non-cool-core clusters may be compared with the ACCEPT data points for multiphase and single-phase clusters, respectively, because the root populations are very similar.  Cool-core clusters tend to reside near the precipitation limit, with an $L_X$--$T$ slope similar to the model prediction, while the non-cool-core clusters reside below it, with a steeper $L_X$--$T$ slope.  Apparently, the processes that shut off production of multiphase gas have a greater impact on core structure in galaxy clusters with shallower potential wells.

Galaxy clusters with central cooling time $> 1$~Gyr generally do not contain multiphase gas in their cores, and this trend appears to extend down through the ATLAS$^{\rm 3D}$ galaxies.  A dotted teal line in Figure~\ref{fig-LX_T} shows the $L_X(R_e)$--$T$ relation corresponding to ambient gas with a constant cooling time $t_{\rm cool} = 1$~Gyr within $R_e = (8.8 \, {\rm kpc}) (kT / {\rm 1 \, kpc})^{0.9}$.  Systems below this line must have central cooling times exceeding 1~Gyr, and the line runs close to the transition from single-phase to multiphase ACCEPT cluster cores.  At lower temperatures, all of the multiphase systems lie above this line, but some of the single-phase systems lie below it.  This pattern may represent another link between the feedback mechanism in galaxy clusters and the mechanisms that regulate early-type galaxies of lower mass.

\section{Summary}	
\label{sec-summary}

This paper has compiled a broad array of $L_X$--$T$ data measured within a variety of apertures and has compared it with a simple model based on the precipitation framework for AGN feedback.  Its main findings are:
\begin{enumerate}

\item The maximum measured values of $L_X(R)$ track the theoretical precipitation limit derived from the condition $t_{\rm cool} \gtrsim 10 t_{\rm ff}$ over nearly seven orders of magnitude in $L_X$, from the cores of massive galaxy clusters all the way down to the Milky Way.  This finding strongly suggests that a common mechanism regulates cooling of circumgalactic gas and possibly also quenching of star formation in all of these massive galactic systems.

\item When measured within the stellar envelopes of early-type galaxies, the $L_X$--$T$ relation changes slope from $\appropto T^{4.5}$ at $kT \lesssim 1$~keV to $\appropto T^{3}$ at $kT \gtrsim 2$~keV.  The change in slope accords with the predictions of precipitation-regulated feedback models, in which $L_X(R) \propto T^3 \Lambda^{-1}(T) R$.  In this context, the bend observed at $\sim 1$~keV is an inversion of the bend in $\Lambda(T)$ in this same temperature range.

\item The $L_X$--$T$ relation measured within the cosmological aperture $R_{500}$ shows little change in slope over the entire observed temperature range.  We construct a model for the $L_X(R_{500})$--$T$ relation by combining a precipitation-regulated entropy profile at small radii with a cosmological entropy profile at large radii and show that the resulting relation also shows little change in slope.

\end{enumerate} 

Observations of X-ray luminosity out to $\sim R_{500}$ around individual massive galaxies may become possible with the next generation of X-ray telescopes and will have much to teach us about the astrophysics of AGN feedback and star-formation quenching.  The large gap between our model for $L_X(R_{500})$--$T$ and measurements of $L_X(R_e)$ in the 0.2--0.7~keV temperature range suggests that such observations will be fruitful.  In this context, it is notable that both the \citet{Anderson_2015MNRAS.449.3806A} points and the Milky Way points at 100~kpc are consistent with our $L_X(R_{500})$--$T$ model for a median $t_{\rm cool}/t_{\rm ff} \approx 20$--30, since this is the same median that best describes the $L_X(R)$--$T$ relation measured within the stellar envelopes of early-type galaxies containing multiphase gas.  Quite possibly, the results of AGN feedback might turn out to be much more systematic and predictable than one would guess from watching movies of numerical simulations.

\vspace*{2em}

We would like to thank Joel Bregman, Greg Bryan, Ali Crocker, Gus Evrard, Mike McCourt, Brian McNamara, Brian O'Shea, and Prateek Sharma for helpful conversations.  This research was supported in part by the National Science Foundation under Grant No. NSF PHY-1125915, because GMV benefitted from the Kavli Institute for Theoretical Physics while participating in their workshop on the Galaxy-Halo Connection.  CPM acknowledges grants NSF AST-1411945 and 1411642.  JG also acknowledges NSF AST-1411642.  MS would like to thank NASA for support through Chandra Award Number AR7-18016X issued by the Chandra X-ray Observatory Center, which is operated by the Smithsonian Astrophysical Observatory for and on behalf of the National Aeronautics Space Administration under contract NAS8-03060.

\bibliographystyle{apj}

\end{document}